\documentclass[a4paper,11pt]{article}
\usepackage[utf8]{inputenc}
\usepackage[english]{babel}
\usepackage{braket}
\usepackage{xspace}
\usepackage{bbm}
\usepackage{physics}
\usepackage{mathdots}
\usepackage{amsfonts}
\usepackage{stmaryrd}
\usepackage{stackrel}
\usepackage[dvipsnames]{xcolor}
\usepackage{appendix}
\usepackage{hyperref}
\usepackage{tikz}
\usepackage{pgfplots}
\usepackage{amsmath}
\usepackage[dvipsnames]{xcolor}
\usepackage{float} % Indispensable pour le [H]
\hypersetup{
 colorlinks=true,
 linkcolor=blue,
 anchorcolor = blue,
 citecolor = blue,
 filecolor = blue,
 urlcolor = blue
}
\usepackage{mathtools}
\usepackage{bbm}
\usepackage{comment}
\usepackage{subfigure}
\usepackage[T1]{fontenc}               % codifica dei font
\usepackage[left=1.8cm,
            right=1.8cm,
            top=2.5cm,
            bottom=3cm
            ]{geometry}                % spaziatura bordi
\usepackage{graphicx}
\usepackage{mathrsfs}
\usepackage{appendix}
\usepackage{fancyhdr}
\usepackage{amsmath,                   % simboli matematici
            amssymb,                   % altri simboli matematici
            amsthm}                    % stili teorema

\usepackage{setspace}
\usepackage{cite}
\usepackage{cancel}
\usepackage[margin=30pt, bf, font=small, center, justification=justified]{caption}[2004/07/16]

\usepackage{setspace}

\title{\bf Entanglement Asymmetry and the Quantum Mpemba Effect in Inhomogeneous Quenches}

\author{Maëlle Olivier$^{1,2}$, Filiberto Ares$^{1, 3}$ and Pasquale Calabrese$^{1}$}

\date{}

\begin{document} 

\maketitle
{\small
\vspace{-5mm}  \ \\
{$^{1}$} SISSA and INFN, via Bonomea 265, 34136 Trieste, Italy\\[+0.1cm]
\medskip
{$^{2}$} Université Paris-Saclay, ENS Paris-Saclay, DER de Physique, 91190 Gif-sur-Yvette, France \\[-0.1cm]
\medskip
{$^{3}$} Departamento de Física Teórica and CAPA, Universidad de Zaragoza, Pedro Cerbuna 12, 50009 Zaragoza, Spain \\[-0.1cm]
}

\begin{abstract}

Entanglement asymmetry is a powerful probe of symmetry breaking and its dynamical restoration following quantum quenches. While extensively studied for homogeneous initial states, its evolution from inhomogeneous initial configurations remains largely unexplored. We investigate a spin-$1/2$ XX chain whose two halves are initially prepared in tilted ferromagnetic states with different tilt angles. Combining generalized hydrodynamics (GHD) and the quasiparticle picture with exact fermionic correlation-matrix calculations, we study the dynamics of entanglement asymmetry. We find good agreement between the two approaches for generic initial configurations and observe the quantum Mpemba effect, whereby states with stronger initial symmetry breaking can restore the symmetry faster. This effect now depends on the subsystem position: while the homogeneous behavior is recovered when the subsystem lies entirely within one of the initially tilted halves, the entanglement-asymmetry curves for subsystems straddling the interface can exhibit multiple crossings or no crossings at all. A notable exception occurs for subsystems entirely within an initially non-tilted half. In this case, GHD predicts vanishing asymmetry, but the exact dynamics generates a small, nonzero asymmetry that can develop a plateau before decaying toward zero. This feature reveals contributions beyond the leading-order hydrodynamic description.

\end{abstract}

\tableofcontents

\section{Introduction}

An important question in the study of nonequilibrium quantum dynamics concerns the fate of symmetries that are broken by the initial state but preserved by the subsequent time evolution. Although unitary evolution cannot restore such symmetries at the level of the global state, they may be dynamically restored within finite subsystems. Early studies of this phenomenon primarily focused on local
observables~\cite{fcec-13,ef-16,ce-20}. More recently, however,
entanglement asymmetry has emerged as a powerful and universal
probe for characterizing its occurrence and understanding the
underlying physical mechanisms~\cite{entanglement_asymmetry}. Defined as the difference between the von Neumann entropies of a reduced density matrix and its symmetrized counterpart, it quantifies the degree of symmetry breaking within a subsystem. In particular, the entanglement asymmetry vanishes if and only if the reduced density matrix respects the symmetry, making it a natural probe of its possible local restoration. 

The dynamics of the entanglement asymmetry has been investigated in several quantum quench protocols, revealing a rich phenomenology. In particular, considerable attention has been devoted to homogeneous symmetry-breaking quenches, in which the system is initially prepared in a spatially uniform state that breaks a symmetry. A paradigmatic example is a spin chain initialized in a tilted ferromagnetic state, namely a product state in which all spins are tilted by the same angle with respect to the $z$-axis~\cite{entanglement_asymmetry}. This state breaks the $U(1)$ symmetry associated with rotations about the $z$-axis, corresponding to particle-number symmetry in the fermionic representation. For integrable quenches, the dynamics of the entanglement asymmetry can be understood within a quasiparticle picture, in which pairs of ballistically propagating quasiparticles carry the pairing correlations responsible for the initial symmetry breaking~\cite{entanglement_asymmetry, rkacmb-24, xy, carc-24, bkccr-24, rvc-24, klobas-24,tc-26,vca-26,ccr-26}. A particularly striking phenomenon emerging from this dynamics is the quantum Mpemba effect~\cite{mpemba1969, bechhoefer-21, qme, teza-26, ylz-25, calabrese-26}, whereby states with stronger initial symmetry breaking can exhibit faster local symmetry restoration than those with weaker symmetry breaking. The quasiparticle picture provides a microscopic explanation of this phenomenon~\cite{rkacmb-24}: it occurs when, in the initially more asymmetric state, the dominant contributions to symmetry breaking are carried by the fastest  quasiparticle pairs generated by the quench. Remarkably, the quantum Mpemba effect has also been found in nonintegrable systems~\cite{liu-24, tcdl-25, fcb-25, yu-25, li-26, amt-26, arst-26, yac-26, yjzxf-25, ya-26, mpf-26, summer-26, lzy-25,amcp-25}, where the quasiparticle picture does not generally apply. Furthermore, the dynamics of entanglement asymmetry has been experimentally measured in quantum simulators, with some of these experiments also reporting the quantum Mpemba effect~\cite{exp_ti,xu-26,yjahzc-26}.

Despite these advances, the dynamics of entanglement asymmetry for spatially inhomogeneous initial states remains much less understood. This raises the question of how such inhomogeneities affect the possible local restoration of symmetry and the occurrence of the quantum Mpemba effect. In this work, we address this question by considering an inhomogeneous quench in a spin-$1/2$ XX chain, where the left and right halves are initially prepared in tilted ferromagnetic states with different tilt angles $\theta_1$ and $\theta_2$.

Generalized hydrodynamics (GHD) provides a systematic framework for describing the large-scale dynamics of integrable systems with spatially inhomogeneous initial conditions~\cite{bertini2016, cady-16} (see also the reviews~\cite{bbdv-22, essler-22, doyon-20, abfpr-21,  dgmsv-25}). Within this framework, the system is locally characterized by a momentum-dependent quasiparticle occupation function, whose  evolution is governed by an Euler hydrodynamic equation. From the resulting occupation profiles, one can determine the evolution of local conserved quantities and their associated currents. Beyond these observables, GHD can also be combined with the quasiparticle picture of entanglement spreading~\cite{cc-05, fc-08, ac-18, ac-17, calabrese-20} to study the entanglement dynamics of inhomogeneous systems~\cite{a-18,abfpr-21, bertini18, abf-19, tckb-26,sa-26}.

Building on these developments, we combine GHD with the quasiparticle picture of entanglement asymmetry to describe its evolution in spatially inhomogeneous quenches. We derive hydrodynamic expressions for the R\'enyi entanglement entropies and charged moments, from which we obtain the entanglement asymmetry, and compare our predictions with exact numerical results from the time-evolved fermionic correlation matrices. This allows us to investigate how the initial tilt angles and the position of the subsystem influence local symmetry restoration and the possible occurrence of the quantum Mpemba effect.

The remainder of the paper is organized as follows. We first introduce in Sec.~\ref{sec:setup} the quench protocol and define the observables of interest, in particular the entanglement asymmetry. In Sec.~\ref{sec:ghd}, we describe the time evolution in terms of fermionic two-point correlation matrices and introduce the GHD framework, which characterizes the large-scale dynamics through the quasiparticle occupation function. Next, in Sec.~\ref{sec:entanglement_asymmetry}, we investigate the dynamics of R\'enyi entanglement entropies and entanglement asymmetry, comparing the GHD predictions with the exact results obtained from the correlation matrices. Finally, in Sec.~\ref{sec:conclusions}, we summarize our findings and discuss possible directions for future work. We also include an appendix with detailed derivations of some analytical expressions presented in the main text.

\section{Quench setup}
\label{sec:setup}

This section lays out the quench protocol and defines the primary physical observables studied throughout this work.

\begin{figure}[t]
\centering
\includegraphics[width=0.9\textwidth]{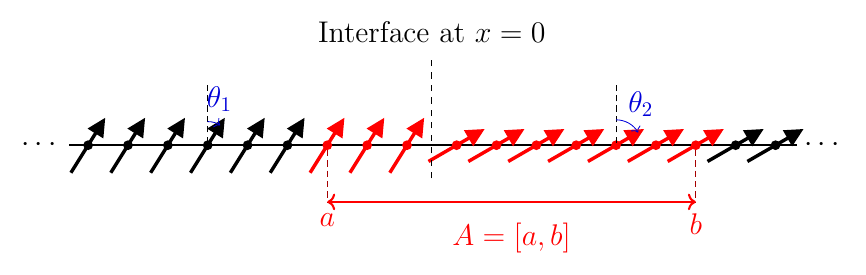}
\caption{
Schematic representation of the initial state $\ket{\psi_0} = \ket{\theta_1\cdots\theta_1\,\theta_2\cdots\theta_2}$. The angles $\theta_1, \theta_2$ denote the spin orientations in the $xz$-plane, measured from the positive $z$-axis, with $\ket{\theta_i=0}=\ket{\uparrow}$, $\ket{\theta_i=\pi}=\ket{\downarrow}$, and $\ket{\theta_i=\frac{\pi}{2}}=\ket{\rightarrow}$ ($i \in \{1,2\}$). The system subsequently evolves unitarily under the tight-binding Hamiltonian~\eqref{eq:hamiltonian}. We consider an infinite chain with lattice sites $j \in \mathbb{Z}$. The subsystem of interest is $A = [a,b]$ (in red), with $N_A=b-a+1$ sites and length $\ell = b - a$ in the continuum.
}
\label{fig:initial_state_scheme}
\end{figure}

The quench protocol is schematically illustrated in Fig.~\ref{fig:initial_state_scheme}. Specifically, we prepare the left and right halves of an infinite spin-$1/2$ chain in product states characterized by different tilting angles, $\theta_1$ and $\theta_2$, respectively, with respect to the $z$-axis. The resulting initial state is given by
\begin{equation}\label{eq:initial_state}
    \ket{\psi_0} = \ket{\cdots \theta_1\cdots\theta_1\,\theta_2\cdots\theta_2\cdots}.
\end{equation}
The subsequent unitary evolution is governed by the XX Hamiltonian, which, upon applying the Jordan--Wigner transformation, can be expressed in terms of spinless fermions as
\begin{equation}
H = -\frac{1}{2} \sum_{j=-\infty}^\infty \left( c_{j+1}^\dagger c_j + c_j^\dagger c_{j+1} \right),
\label{eq:hamiltonian}
\end{equation}
where $c_j, c_j^\dagger$ are spinless annihilation and creation fermionic operators at the $j$-th site. 
%We impose periodic boundary conditions, $c_{L+1}=c_1$, on the fermionic Hamiltonian. %The specific choice of boundary conditions is not relevant in the regime considered here, since GHD is formulated in the thermodynamic limit $L\to \infty$  and the numerical analysis is restricted to times before finite-size effects become relevant.

The initial state~\eqref{eq:initial_state} explicitly breaks the $U(1)$ symmetry associated with rotations about the $z$-axis whenever at least one of the tilt angles differs from $0$ and $\pi$. In fermionic language, these rotations are generated by the charge $Q=\sum_{j=-\infty}^{\infty}c_j^\dagger c_j$. By contrast, the Hamiltonian commutes with $Q$ and thus preserves the corresponding $U(1)$ symmetry. 
Our main goal is to study the evolution of this symmetry within a subsystem and its possible local restoration at long times. The degree of symmetry breaking within the subsystem can be quantified by the entanglement asymmetry~\cite{entanglement_asymmetry}. The evolution of the entanglement asymmetry under the Hamiltonian~\eqref{eq:hamiltonian} has been studied in the homogeneous case $\theta_1=\theta_2$ in Ref.~\cite{entanglement_asymmetry}. This analysis was later extended to generic fermionic squeezed initial states~\cite{rkacmb-24, xy, carc-24} and to two dimensions~\cite{yac-24}. 

To define the entanglement asymmetry, we consider a subsystem $A$ of the chain. Throughout this work, the lattice subsystem contains both endpoints and consists of $N_A=b-a+1$ sites and, in the continuum hydrodynamic description, we denote the interval length by $\ell=b-a$. Its reduced density matrix is given by $\rho_A(t) = \mathrm{Tr}_B \left( \ket{\psi(t)} \bra{\psi(t)} \right)$, where $B$ denotes the complement of $A$. The $U(1)$ symmetry is preserved within the subsystem if and only if $\rho_A$ commutes with the restricted charge operator $Q_A = \sum_{j \in A} c_j^\dagger c_j$. Equivalently, $\rho_A$ must be block diagonal in the eigenspaces of $Q_A$. When the symmetry is broken, $\rho_A$ contains off-diagonal blocks connecting different charge sectors. These off-diagonal blocks can be removed by introducing the symmetrized reduced density matrix
\begin{equation}
\rho_{A,Q} = \sum_{q \in \mathbb{Z}} \Pi_q \rho_A \Pi_q,
\end{equation}
where $\Pi_q$ is the projector onto the eigenspace of $Q_A$ with eigenvalue $q$. By construction, $\rho_{A,Q}$ commutes with $Q_A$ and coincides with $\rho_A$ when the latter preserves the symmetry. The entanglement asymmetry is then defined as the difference between the von Neumann entropies of $\rho_{A,Q}$ and $\rho_A$:
\begin{equation}
\Delta S_A(t) = S(\rho_{A,Q}(t)) - S(\rho_{A}(t)), \qquad S(\rho) = - \mathrm{Tr}(\rho \log\rho).
\label{eq:entanglement_asymmetry_def}
\end{equation}
Equivalently, the entanglement asymmetry can be expressed as the relative entropy between $\rho_A$ and $\rho_{A,Q}$,
\begin{equation}\label{eq:entanglement_asymmetry_def_rel_ent}
\Delta S_A(t) = \mathrm{Tr}\left(\rho_A(t)(\log\rho_A(t) - \log\rho_{A,Q}(t))\right).
\end{equation}
The positivity of the relative entropy guarantees that $\Delta S_A(t) \geq 0$, with equality if and only if $\rho_A(t) = \rho_{A,Q}(t)$, corresponding to full local symmetry restoration.

In practice, Eqs.~\eqref{eq:entanglement_asymmetry_def} and~\eqref{eq:entanglement_asymmetry_def_rel_ent} are very difficult to evaluate due to the presence of the logarithm. Alternatively, the degree of symmetry breaking can be analyzed through the R\'enyi entanglement asymmetry,
\begin{equation}
\Delta S_A^{(n)}(t) = S^{(n)}(\rho_{A,Q}(t)) - S^{(n)}(\rho_A(t)),
\label{eq:renyi_moments}
\end{equation}
where the von Neumann entropy is replaced by the R\'enyi entropy $S^{(n)}(\rho)= 1/(1-n) \log\mathrm{Tr}(\rho^n)$.
The R\'enyi entanglement asymmetries satisfy $\Delta S^{(n)}_A \ge 0$. We recover the von Neumann asymmetry~\eqref{eq:entanglement_asymmetry_def} in the limit $n \to 1$. For integer $n$, the R\'enyi entanglement asymmetries have the advantage of being more readily accessible analytically and numerically, and can also be measured experimentally in quantum simulators~\cite{exp_ti,xu-26, yjahzc-26}.

In fact, using the Fourier representation of the projectors $\Pi_q$, the symmetrized reduced density matrix can be expressed as
\begin{equation}
\rho_{A,Q} = \int_{-\pi}^{\pi} \frac{\mathrm{d}\alpha}{2\pi} e^{-i\alpha Q_A}\rho_A e^{i\alpha Q_A}.
\end{equation}
This representation allows the R\'enyi entanglement asymmetries to be computed in terms of the charged moments~\cite{entanglement_asymmetry},
\begin{equation}\label{eq:Znalpha}
Z_n(\vec{\alpha},t) = \mathrm{Tr}\left[\prod_{j=1}^n \rho_A(t)e^{i\alpha_{j,j+1}Q_A}\right],
\end{equation}
where $\vec{\alpha}=(\alpha_1,\dots,\alpha_n)$, $\alpha_{i,j} = \alpha_i-\alpha_j$, and $\alpha_{n+1}=\alpha_1$. In particular, one obtains
\begin{equation}
\Delta S_A^{(n)}(t) = \frac{1}{1-n} \log \left( \int_{-\pi}^{\pi} \dots \int_{-\pi}^{\pi} \frac{\mathrm{d}\alpha_1 \cdots \mathrm{d}\alpha_n}{(2\pi)^n} \frac{Z_n(\vec{\alpha},t)}{Z_n(\vec{0}, t)} \right).
\label{eq:entanglement_asymmetry}
\end{equation}
The R\'enyi entropy is recovered in the case $\alpha_1=\cdots\alpha_n=0$, 
\begin{equation}\label{eq:renyi_entropy}
    S_A^{(n)} = \frac{1}{1-n} \log Z_n(\vec{0}).
\end{equation}

Although the initial spin-product state~\eqref{eq:initial_state} is not Gaussian in the fermionic representation, we follow Ref.~\cite{entanglement_asymmetry} and consider instead its Gaussian cat-state counterpart, obtained by projecting onto a fixed fermion-parity sector (see Refs.~\cite{entanglement_asymmetry,xy} for all the technical details). Since the Hamiltonian~\eqref{eq:hamiltonian} is quadratic, Gaussianity is preserved throughout the time evolution. Consequently, the time-evolved cat state is fully characterized by the two-point correlations, $C_{j,j'}=\langle c_j^\dagger c_{j'}\rangle$ and $F_{j,j'}=\langle c_j c_{j'}\rangle$. 
Throughout the following, the exact numerical results refer to this Gaussian cat-state representation.
For a subsystem $A$ of $N_A$ sites, these correlations can be assembled into a full covariance matrix~\cite{peschel2003,pe-09}
\begin{equation}
\Gamma_A(t)=
\begin{pmatrix}
I_{N_A}-2C_A^\dagger(t) & 2F_A(t)\\
2F_A^\dagger(t) & 2C_A(t)-I_{N_A}
\end{pmatrix},
\label{eq:corr_matrix}
\end{equation}
of dimension $2N_A\times 2N_A$. Here, $C_A(t)$ and $F_A(t)$ denote the restriction of $C_{j, j'}$ and $F_{j, j'}$ to the sites in $A$.
The charged moments in Eq.~\eqref{eq:Znalpha} can be then explicitly expressed as~\cite{entanglement_asymmetry, amvc-23}:
\begin{equation}
Z_n(\vec{\alpha}, t) = \sqrt{ \det \left[ \left( \frac{I_{2N_A}-\Gamma_A(t)}{2} \right)^n \left( I_{2N_A} + \prod_{j=1}^{n} \frac{I_{2N_A}+\Gamma_A(t)}{I_{2N_A}-\Gamma_A(t)} e^{i\alpha_{j,j+1}n_A} \right) \right] },
\label{eq:z_n}
\end{equation}
where $n_A = \text{diag}(-I_{N_A}, I_{N_A})$ is a $2N_A\times2N_A$ diagonal matrix (notice that, as usual, the eigenvalues of $\Gamma_A$ equal to $1$ do not cause any singularity in the complete determinant expression because they cancel with the prefactor having  zeros of the same order exactly in $1$). In particular, for $n=2$, Eq.~\eqref{eq:z_n} simplifies to
\begin{equation}
Z_2(\alpha,t) = \sqrt{ \det \left( \frac{ I_{2N_A}+\Gamma_{\alpha}(t)\Gamma_{-\alpha}(t) }{2} \right) }, \qquad \Gamma_{\alpha}(t) = \Gamma_A(t)\,e^{i\alpha n_A}.
\label{eq:z2}
\end{equation}
This determinant representation is particularly convenient, as it allows the charged moments to be computed directly from the two-point correlation functions, without explicitly constructing the reduced density matrix. In particular, it reduces the calculation of the entanglement asymmetry to the evaluation of determinants of matrices whose dimensions scale linearly with the subsystem size, providing an efficient framework for both analytical and numerical investigations of its time evolution.

\section{Time evolution of correlation matrices and occupation functions}
\label{sec:ghd}

In this section, we determine the time evolution of the two-point correlation functions following our quench protocol, which allows us to compute the entanglement asymmetry numerically using the formalism introduced in Sec.~\ref{sec:setup}. We also employ generalized hydrodynamics (GHD) to determine the time evolution of the mode occupation densities, which provides the basis for deriving analytical expressions for the entanglement asymmetry in the following section.

At $t=0$, the two-point correlation matrices $C(t)$ and $F(t)$ have a block structure  reflecting the preparation of the left and right halves of the chain in two distinct homogeneous states:
\begin{equation}
C(0)= \begin{pmatrix} C^{(11)} & C^{(12)}\\ C^{(21)} & C^{(22)} \end{pmatrix}, \qquad F(0)= \begin{pmatrix} F^{(11)} & F^{(12)}\\ F^{(21)} & F^{(22)} \end{pmatrix}.
\end{equation}
Here, the superscripts $1$ and $2$ refer to the left and right halves, respectively, with site indices $j\leq 0$ in the left half and $j\geq 1$ in the right half. The diagonal blocks describe correlations between sites belonging to the same half of the chain. For $m=1,2$, their matrix elements are given by
\begin{equation}
C^{(mm)}_{j,j'}= \begin{cases} \cos^2(\theta_m/2), & j=j', \\[0.1cm] \frac{1}{4}\sin^2(\theta_m)(-\cos\theta_m)^{|j-j'|-1}, & j\neq j', \end{cases} \quad
F^{(mm)}_{j,j'}= \begin{cases} 0, & j=j', \\[0.1cm] -\frac{1}{4}\sin^2(\theta_m)(-\cos\theta_m)^{j-j'-1}, & j>j', \end{cases}
\label{eq:c_diag_blocks}
\end{equation}
with the remaining elements of $F^{(mm)}$ determined by antisymmetry, $F^{(mm)}_{j,j'}=-F^{(mm)}_{j',j}$.
The off-diagonal blocks, instead, encode correlations between sites in opposite halves of the chain, with $j\leq 0$ and $j'\geq 1$, and satisfy
\begin{equation}
C^{(12)}_{j,j'} = C^{(21)}_{j',j} = F^{(12)}_{j,j'} = -F^{(21)}_{j',j} = \frac{1}{4}\sin(\theta_1)\sin(\theta_2)(-\cos\theta_1)^{-j}(-\cos\theta_2)^{j'-1}.
\label{eq:c_cross_blocks}
\end{equation}
These correlations are $2\pi$-periodic in each tilt angle and are invariant under the simultaneous transformation $(\theta_1,\theta_2)\mapsto(-\theta_1,-\theta_2)$. In the following, we restrict to $\theta_1,\theta_2\in[0,\pi]$.

Having determined the initial correlation matrices $C(0)$ and $F(0)$, we now turn to their exact time evolution. By diagonalizing the quadratic Hamiltonian~\eqref{eq:hamiltonian} through a spatial Fourier transform, their dynamics can be expressed in terms of the single-particle propagator $J(t)$ as~\cite{pe-09}
\begin{equation}
C(t) = J^\dagger(t)\,C(0)\,J(t), \qquad F(t) = J^{T}(t)\,F(0)\,J(t).
\label{eq:corr_evolutions}
\end{equation}
For the infinite chain, the matrix elements of the propagator take the explicit form
\begin{equation}
J_{j,j'}(t) = \int_{0}^{2\pi} \frac{{\rm d}k}{2\pi} e^{i t \cos k - ik(j-j')} = i^{j-j'} \mathcal{J}_{j-j'}(t),
\label{eq:propagator_bessel}
\end{equation}
where $\mathcal{J}_n(t)$ denotes the Bessel function of the first kind. Since we are considering an infinite chain, the correlation matrices $C(t)$ and $F(t)$, as well as the propagator $J(t)$, are infinite-dimensional. For numerical calculations, we truncate these matrices to a finite interval of $L$ sites centered at the junction. The cutoff $L$ is chosen sufficiently large to avoid truncation effects over the time intervals considered.

While the exact evolution of the correlation matrices provides a complete microscopic description of the dynamics, the large-scale behavior of the quench protocol depicted in Fig.~\ref{fig:initial_state_scheme} can be understood within GHD~\cite{bertini2016, cady-16}. Integrable quantum many-body systems possess an infinite set of local conserved charges, which allow the local state at the Euler scale to be characterized by a quasiparticle occupation function $n(k,x,t)$. This function describes the occupation of quasiparticles with momentum $k$ at position $x$ and determines the local densities and currents of conserved quantities. In our free-fermion model, quasiparticles propagate ballistically with momentum-dependent velocities, and the spatial inhomogeneity of the initial state is encoded in the position dependence of the occupation function. In the hydrodynamic limit, this function can be defined through the Wigner transform of the fermionic two-point correlation function~\cite{wigner-32,hpb-12}
\begin{equation}
n(k, x, t)=\int {\rm d}z \,{\rm Tr}\left(\rho(t)c_{x+z/2}^\dagger c_{x-z/2}\right)e^{ikz},
\end{equation}
where the site index $j$ is replaced by the continuum variable $x$.

In our case, since the Hamiltonian is free and homogeneous, the Euler-scale dynamics is governed by the transport equation~\cite{bertini2016, cady-16, bertini18}
\begin{equation}
    \bigl(\partial_t + v(k)\,\partial_x\bigr)n(k,x,t)=0.
    \label{eq:euler}
\end{equation}
Here, $v(k)=\sin k$ is the group velocity associated with the single-particle dispersion relation of the XX Hamiltonian~\eqref{eq:hamiltonian}. The solution to Eq.~\eqref{eq:euler} is given by 
\begin{equation}
    n(k,x,t) = n \bigl(k,x-v(k)t,0\bigr).
    \label{eq:n_k_t}
\end{equation}
Therefore, for the specific inhomogeneous quench geometry shown in Fig.~\ref{fig:initial_state_scheme}, the time-evolved occupation function~\eqref{eq:n_k_t} takes the piecewise form
\begin{equation}
    n(k,x,t) =
    \begin{cases}
        \dfrac{1+\cos\Delta_k(\theta_1)}{2}, & x-t\sin(k)\leq 0, \\[0.2cm]
        \dfrac{1+\cos\Delta_k(\theta_2)}{2}, & x-t\sin(k)>0,
    \end{cases}
    \label{eq:expression_nk}
\end{equation}
where the functions $\cos\Delta_k(\theta_i)$ encode the initial occupations in the two halves via
\begin{equation}
    \cos\Delta_k(\theta_i) = \frac{h(\theta_i)+\cos k}{\sqrt{(h(\theta_i)+\cos k)^2+(\gamma(\theta_i)\sin k)^2}},
\end{equation}
with
\begin{equation}
    h(\theta_i) = \frac{2\cos\theta_i}{1+\cos^2\theta_i}, \qquad \gamma(\theta_i) = \frac{1-\cos^2\theta_i}{1+\cos^2\theta_i}.
\end{equation}

The hydrodynamic evolution~\eqref{eq:n_k_t} also admits a semiclassical quasiparticle interpretation~\cite{cc-05, fc-08, ac-17, ac-18, calabrese-20,rrc-24}, which provides an intuitive picture of the spreading of entanglement and fluctuations. In this picture, the initial state acts as a source of entangled quasiparticle pairs emitted locally from each position $x_0$. Each pair consists of two excitations with opposite momenta $k$ and $-k$, propagating ballistically with group velocities $v(k)$ and $v(-k)=-v(k)$, respectively. Their trajectories are therefore given by $x_\pm(t)=x_0\pm v(k)t$. While quasiparticles emitted from different positions are assumed to be uncorrelated, the two members of each pair remain entangled as they propagate. The distribution of these pairs across the boundaries of a subsystem provides a semiclassical description of the entanglement dynamics and of other quantities such as entanglement asymmetry.

\begin{figure}[t]
    \centering
    \includegraphics[width=\textwidth]{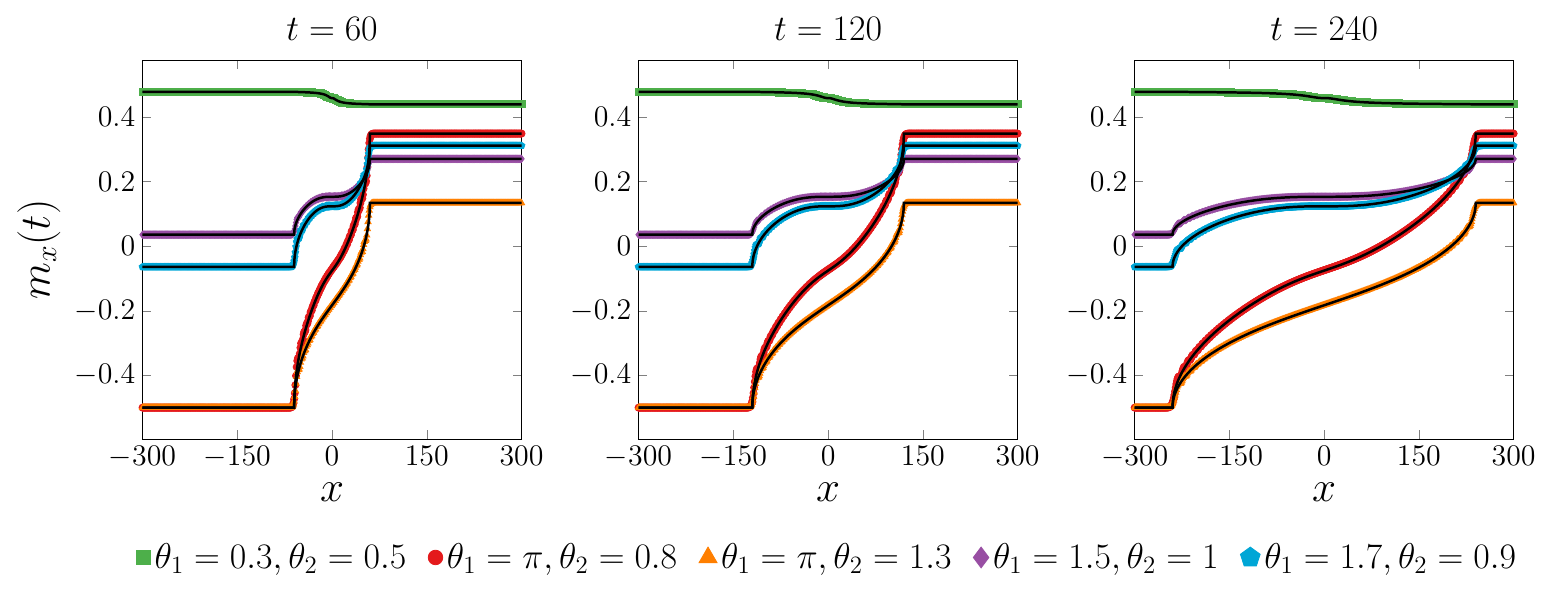}
    \caption{
   Magnetization profiles for different initial tilt angles $\theta_1$ and $\theta_2$ at three times $t$ (from left to right: $60$, $120$, and $240$). Symbols represent the exact local magnetization obtained from the diagonal elements of the time-evolved correlation matrix~\eqref{eq:magnetization_corr}, taking a chain cutoff $L=1400$, while solid lines show the GHD prediction~\eqref{eq:magnetization_ghd}. 
    }
    \label{fig:comparison_magnetization_theta}
\end{figure}

As a first test of the GHD predictions, we compare the time evolution of the local magnetization obtained from the exact correlation matrices~\eqref{eq:corr_evolutions} with that derived from the quasiparticle occupation function~\eqref{eq:n_k_t}. The local magnetization along the $z$-direction is defined as
\begin{equation}
m_j(t) = \frac{1}{2} \langle \psi_t | \hat{\sigma}_j^z | \psi_t \rangle,
\label{eq:magnetization}
\end{equation}
where $\hat{\sigma}_j^z$ denotes the local Pauli operator along the $z$-direction at site $j$, which, under the Jordan–Wigner transformation, is expressed in terms of fermionic operators as $\hat{\sigma}_j^z=2c_j^\dagger c_j-1$. The particular choice $\theta_1=\pi$ and $\theta_2=0$ corresponds to the conventional domain-wall melting problem, whose magnetization dynamics has been extensively studied in the literature~\cite{antal-99, karevski-02, ogata-02, hunyadi-04, pk-07, akr-08, lm-10, er-13, adsv-16, vsdh-16, hydro_scopa}.  In the fermionic representation, the lattice magnetization is directly related to the normal correlation matrix via
\begin{equation}
m_j(t) = -\frac{1}{2} + C_{j,j}(t).
\label{eq:magnetization_corr}
\end{equation}
Concurrently, within the GHD framework, the local magnetization profile in the hydrodynamic limit is obtained by integrating the quasiparticle occupation function over momentum:
\begin{equation}
    m_x(t) = \int_{-\pi}^{\pi} \frac{{\rm d}k}{2\pi}\, n(k,x,t) - \frac{1}{2}.
    \label{eq:magnetization_ghd}
\end{equation}
In Fig.~\ref{fig:comparison_magnetization_theta}, we compare the exact magnetization profiles obtained from Eq.~\eqref{eq:magnetization_corr} (symbols) with the GHD predictions from Eq.~\eqref{eq:magnetization_ghd} (solid curves), for different initial tilt angles and times. We find excellent agreement between the two approaches, confirming the accuracy of the hydrodynamic description of the large-scale magnetization dynamics.

\section{Entanglement entropy and asymmetry}
\label{sec:entanglement_asymmetry}

We now investigate the evolution of entanglement and symmetry breaking within a subsystem using the quasiparticle picture and the hydrodynamic description developed in the previous section. We derive analytical expressions for the R\'enyi entanglement entropies and charged moments, from which we obtain the entanglement asymmetry. Finally, we compare these predictions with exact numerical results obtained from the time-evolved fermionic correlation matrices.

\subsection{Entanglement entropy} 

Let us first focus on the R\'enyi entanglement entropy introduced in Eq.~\eqref{eq:renyi_entropy}. We recall that its time evolution has been extensively studied for the special case $\theta_1=0$ and $\theta_2=\pi$, corresponding to the domain-wall melting problem~\cite{vicari-12, ep-14, ah-14, ge-19, eisler-25, dsvc-17, skcd-21, hydro_scopa}. Here, we extend the analysis to arbitrary tilt angles $\theta_1$ and $\theta_2$.
Within the quasiparticle picture, entanglement entropy originates from entangled quasiparticle pairs shared between the subsystem $A = [a, b]$ and its complement. Consider a quasiparticle pair created at position $y$ with momentum $k$. This pair contributes to the entanglement entropy if and only if one quasiparticle resides in $A$ while the other lies outside. For right-moving quasiparticles ($v(k) > 0$), this condition reads
\begin{equation}
\begin{cases}
a < y + v(k)t < b, \\[0.1cm]
y - v(k)t < a,
\end{cases}
\end{equation}
which simplifies to $x \in \bigl[a,\, \min(a + 2v(k)t,\, b)\bigr]$, with $x=y+v(k)t$. Analogous reasoning applies to left-moving quasiparticles ($v(k) < 0$), yielding a symmetric contribution from the right boundary of the subsystem.
Each shared pair contributes to the Rényi entropy with a weight determined by the Rényi–Yang–Yang entropy density
\begin{equation}
s_{\text{YY}}^{(n)}(u) = \frac{1}{1-n} \log\left(u^n + (1-u)^n\right),
\label{eq:yang_yang}
\end{equation}
evaluated at the local occupation function $u=n(k,x,t)$. Summing over all quasiparticle pairs yields the leading order contribution for the R\'enyi entropy in the hydrodynamic limit~\cite{bertini18}:
\begin{multline}
S_A^{(n)}(t) =\int_0^{2\pi} \frac{\mathrm{d}k}{2\pi} \, \Theta\bigl(-v(k)\bigr) \int_{\max\left(b + 2v(k)t,\, a\right)}^{b} \mathrm{d}x \, s_{\text{YY}}^{(n)}\bigl(n(k,x,t)\bigr) \\
+ \int_0^{2\pi} \frac{\mathrm{d}k}{2\pi} \, \Theta\bigl(v(k)\bigr) \int_{a}^{\min\left(a + 2v(k)t,\, b\right)} \mathrm{d}x \, s_{\text{YY}}^{(n)}\bigl(n(k,x,t)\bigr),
\label{eq:s2_ghd}
\end{multline}
where $\Theta(x)$ is the Heaviside step function.
In the domain-wall melting problem ($\theta_1=0, \theta_2=\pi$), the occupation function $n(k,x,t)$ takes only the values $0$ and $1$, for which the R\'enyi--Yang--Yang entropy density~\eqref{eq:yang_yang} vanishes, $s_{\mathrm{YY}}^{(n)}(0)=s_{\mathrm{YY}}^{(n)}(1)=0$. Consequently, the leading hydrodynamic contribution~\eqref{eq:s2_ghd} to the R\'enyi entanglement entropy is identically zero, and one must consider subleading corrections arising from quantum fluctuations of the occupation function $n(k,x,t)$~\cite{dsvc-17, skcd-21}, which can be captured using quantum generalized hydrodynamics~\cite{rbd-19, rcdd-20, srcd-23, tscvd-24}. By contrast, for generic tilt angles $\theta_1,\theta_2\notin{0,\pi}$, the occupation function $n(k,x,t)$ takes values between $0$ and $1$, leading to a nonvanishing R\'enyi--Yang--Yang entropy density. Consequently, the leading hydrodynamic contribution to the R\'enyi entanglement entropy is nonzero and can be determined within standard GHD, without the need to account for subleading corrections arising from quantum fluctuations. A separate order-one contribution to the R\'enyi entanglement entropy can arise from the Gaussian cat-state structure of the initial state, which is not captured by the leading-order quasiparticle picture. At $t=0$, this contribution approaches $\log 2$ when both the subsystem and its complement contain sufficiently many tilted spins, but vanishes when the subsystem lies entirely within an initially fully polarized region. 

\begin{figure}[t]
    \includegraphics[width = \textwidth]{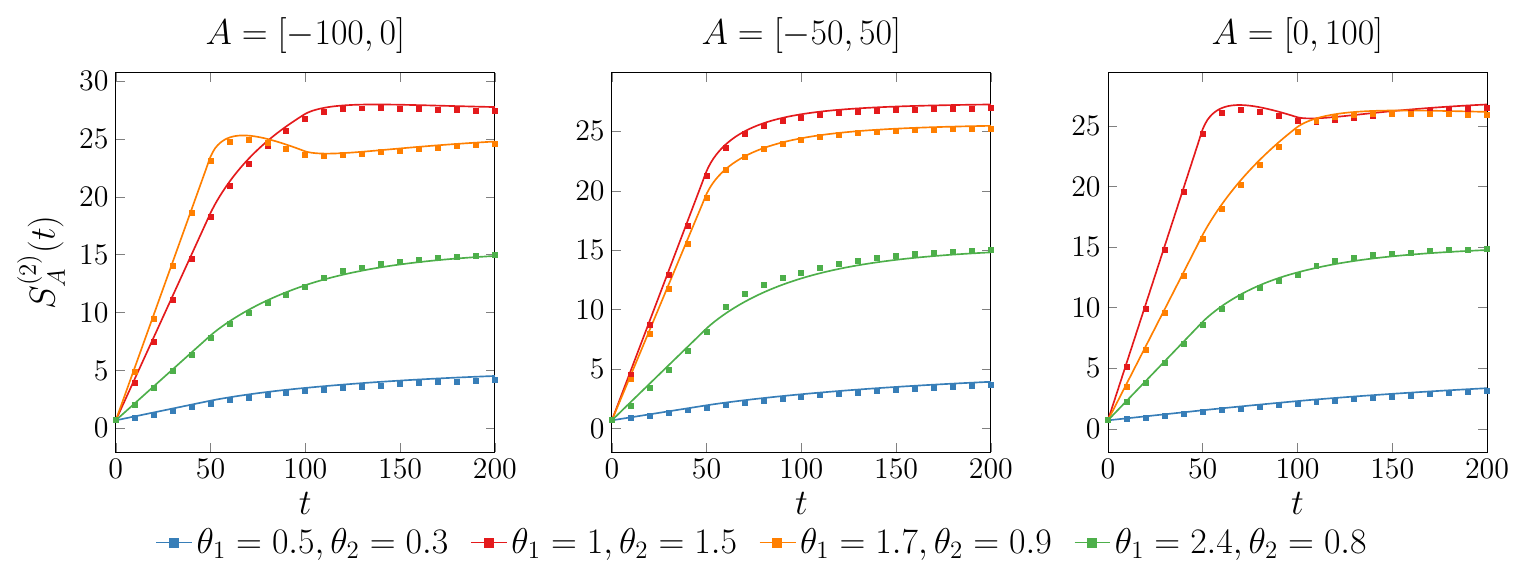}
      \includegraphics[width = \textwidth]{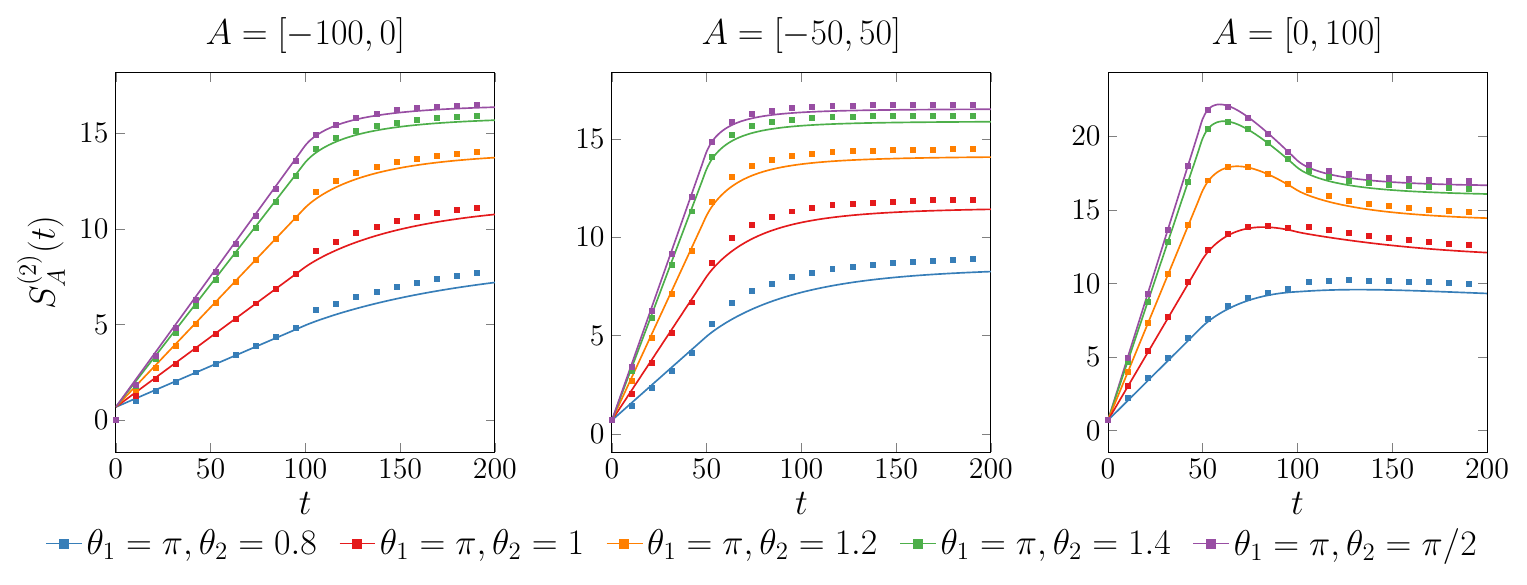}
    \caption{R\'enyi-2 entanglement entropy $S_A^{(2)}(t)$ as a function of time for different initial tilt angles $\theta_1$ and $\theta_2$ and subsystem positions. In the upper row, $\theta_1\neq\theta_2$, with $\theta_1,\theta_2\notin\{0,\pi\}$, while in the lower row, $\theta_1=\pi$, corresponding to the left half of the chain being initially fully polarized down. The left, middle, and right columns correspond to subsystems located entirely within the left half, crossing the interface, and entirely within the right half, respectively. The subsystem length is $\ell=100$ in all panels. Symbols represent the exact results obtained from the correlation matrix~\eqref{eq:corr_evolutions} using Eqs.~\eqref{eq:z2} and \eqref{eq:renyi_entropy}, taking a chain cutoff $L=800$, while solid lines show the GHD predictions~\eqref{eq:s2_ghd}.}
    \label{fig:several_theta_polarized}
\end{figure}

\begin{figure}[t]
    \includegraphics[width = \textwidth]{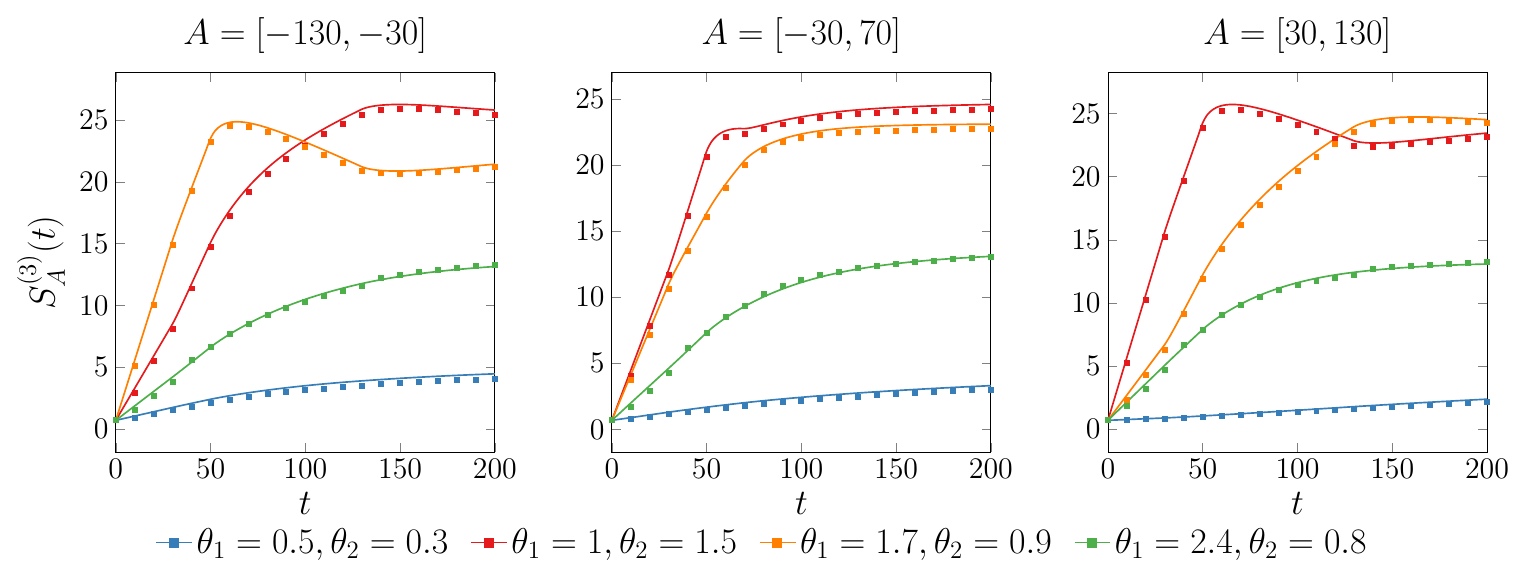}
    \includegraphics[width = \textwidth]{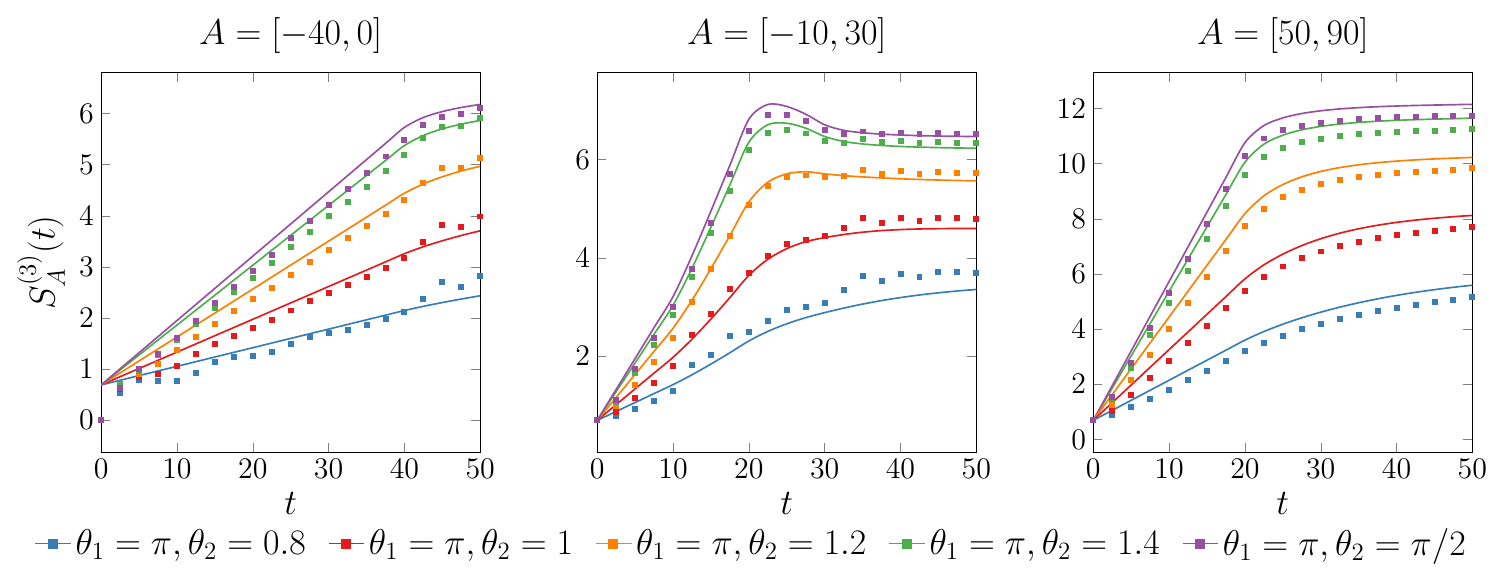}
    \caption{R\'enyi-3 entanglement entropy $S_A^{(3)}(t)$ as a function of time for different initial tilt angles $\theta_1$ and $\theta_2$ and subsystem positions. The upper and lower rows correspond to generic tilt angles and $\theta_1=\pi$, respectively, while the columns distinguish subsystems located to the left of, across, and to the right of the interface. Symbols represent exact correlation-matrix results, taking a chain cutoff $L=800$, while solid lines show the GHD prediction~\eqref{eq:s2_ghd}. The subsystem length is $\ell=100$ in the upper row and $\ell=40$ in the lower row.}
    \label{fig:several_theta3}
\end{figure}

In Figs.~\ref{fig:several_theta_polarized} and~\ref{fig:several_theta3}, we compare the exact R\'enyi-$2$ and R\'enyi-$3$ entropies, respectively, obtained numerically from Eqs.~\eqref{eq:renyi_entropy} and~\eqref{eq:z_n} (symbols) with the GHD prediction from Eq.~\eqref{eq:s2_ghd} (solid curves). Here, they are plotted as a function of time for several initial angles $\theta_1, \theta_2 \in [0, \pi]$. In particular, in the lower row of both figures, $\theta_1$ is fixed to $\pi$, corresponding to an initial state in which the left half of the spin chain is fully polarized down. In both figures, we consider three spatial configurations of the subsystem: entirely within the left half of the chain (left panels), crossing the interface (middle panels), and entirely within the right half (right panels).
%We restrict the time interval to a regime in which finite-size effects associated with the total system size are still negligible. Although these effects could in principle be incorporated into the GHD predictions~\cite{asw-22} and in the quasiparticle picture \cite{mac-20}, they are not considered here. 
Overall, the agreement is excellent. Interestingly, the exact numerical R\'enyi entropies exhibit a sharp change around a characteristic time $t_*$ that is not captured by the GHD predictions. This behavior becomes particularly pronounced as the initial tilt angles approach fully polarized configurations, especially for R\'enyi index $n=3$ and when the left half is fully polarized ($\theta_1=\pi$). For the latter case, the characteristic time depends on the position of the subsystem. When the subsystem straddles the interface, $t_*$ approximately coincides with the number of sites in its left portion ($t_*\approx 50$ in Fig.~\ref{fig:several_theta_polarized} and $t_*\approx 10$ in Fig.~\ref{fig:several_theta3}). By contrast, when the subsystem lies entirely within either half of the chain, the sharp change occurs at $t_*\approx\ell$. The enhancement of this behavior as we approach fully polarized configurations suggests a possible connection with subleading quantum corrections, analogous to those governing entanglement growth in the domain-wall melting problem~\cite{hydro_scopa}.

\subsection{Entanglement asymmetry} 

We now turn to the entanglement asymmetry, whose dynamics in the hydrodynamic regime can also be understood within the quasiparticle picture. This approach was originally developed for homogeneous quenches in Ref.~\cite{entanglement_asymmetry, xy}, where the evolution of the charged moments~\eqref{eq:Znalpha} can be described in terms of the entangled pairs of quasiparticles generated in the quench that propagate ballistically. Here, we extend this framework to the inhomogeneous quench under consideration, incorporating the spatial dependence of the quasiparticle occupations through GHD. 

Following Ref.~\cite{entanglement_asymmetry, xy}, the contribution to the logarithmic ratio $\log[Z_n(\vec{\alpha},t)/Z_n(\vec{0},t)]$ of a mode with occupation $u$ is given by
\begin{equation}
f^{(n)}(u, \vec{\alpha}) = \frac{1}{2} \sum_{j=1}^n \log f\bigl(2u-1, \alpha_j - \alpha_{j+1}\bigr),
\label{eq:f_charged}
\end{equation}
with $f(\lambda, \alpha) = \mathrm{i} \lambda \sin(\alpha) + \cos(\alpha)$. The functions $f^{(n)}$ play a role analogous to the Yang--Yang entropy density $s_{\text{YY}}^{(n)}$. In the present inhomogeneous setting, we assume that this contribution is determined by the local occupation function~\eqref{eq:expression_nk} that evolves according to Eq.~\eqref{eq:n_k_t}.

Unlike the entanglement entropy, the charged moments receive contributions only from quasiparticle pairs whose two members belong to the subsystem $A$ at time $t$~\cite{entanglement_asymmetry, xy}. These pairs can be identified as propagating Cooper pairs, which carry the pairing correlations responsible for breaking the $U(1)$ particle-number symmetry in the initial state. For a pair emitted at position $y$, the condition that both quasiparticles lie within $A=[a,b]$ restricts the emission point to $y\in[a+|v(k)|t,b-|v(k)|t]$. Integrating over all allowed emission positions and quasiparticle momenta, we obtain the following hydrodynamic prediction for the charged moments:
\begin{multline}
\log\left(\frac{Z_n(\vec{\alpha},t)}{Z_n(\vec{0},t)}\right) = \int_0^{2\pi} \frac{\mathrm{d}k}{2\pi} \, \Theta\bigl(-v(k)\bigr)\Theta\bigl(b-a-2|v(k)|t\bigr) \int_{a - v(k)t}^{b + v(k)t} \mathrm{d}y \, f^{(n)}\!\bigl(n(k,y,0), \vec{\alpha}\bigr) \\
+ \int_0^{2\pi} \frac{\mathrm{d}k}{2\pi} \, \Theta\bigl(v(k)\bigr)\Theta\bigl(b-a-2|v(k)|t\bigr) \int_{a + v(k)t}^{b - v(k)t} \mathrm{d}y \, f^{(n)}\!\bigl(n(k,y,0), \vec{\alpha}\bigr).
\label{eq:deltas2_ghd}
\end{multline}
Combined with Eq.~\eqref{eq:entanglement_asymmetry}, Eq.~\eqref{eq:deltas2_ghd} provides the hydrodynamic description of the entanglement asymmetry dynamics.

\begin{figure}[t!]
    \centering
    \includegraphics[width=\textwidth]{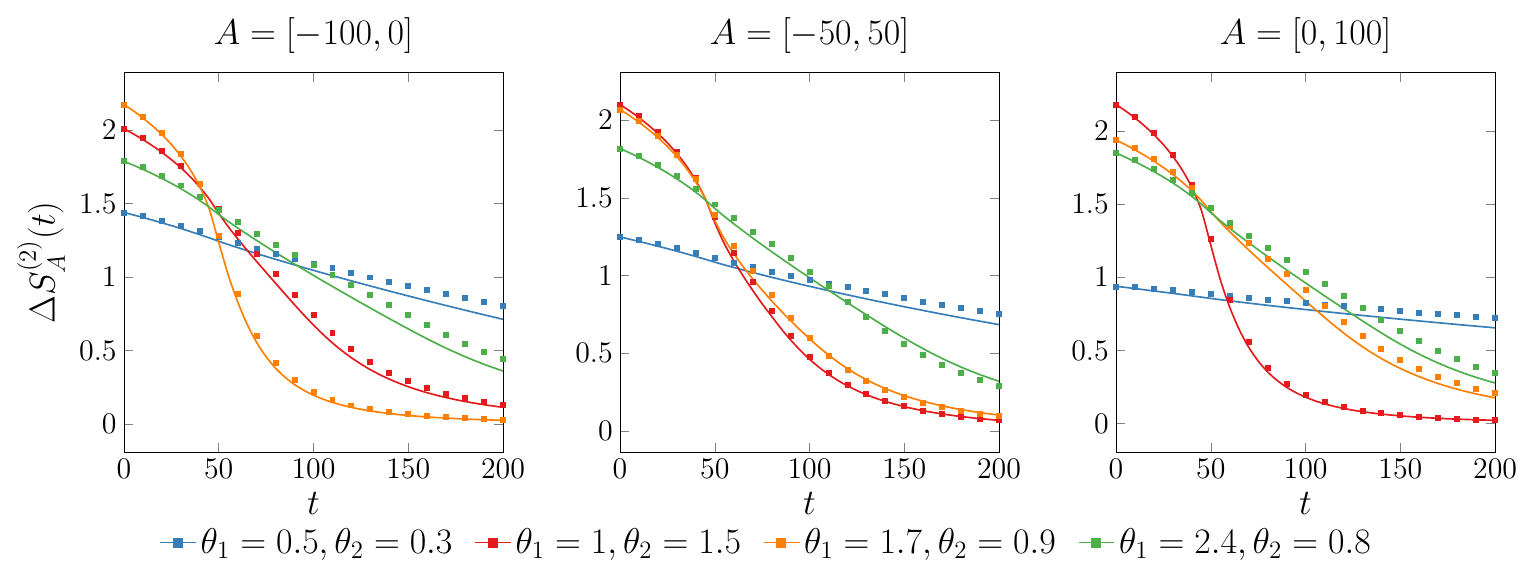}
     \includegraphics[width=\textwidth]{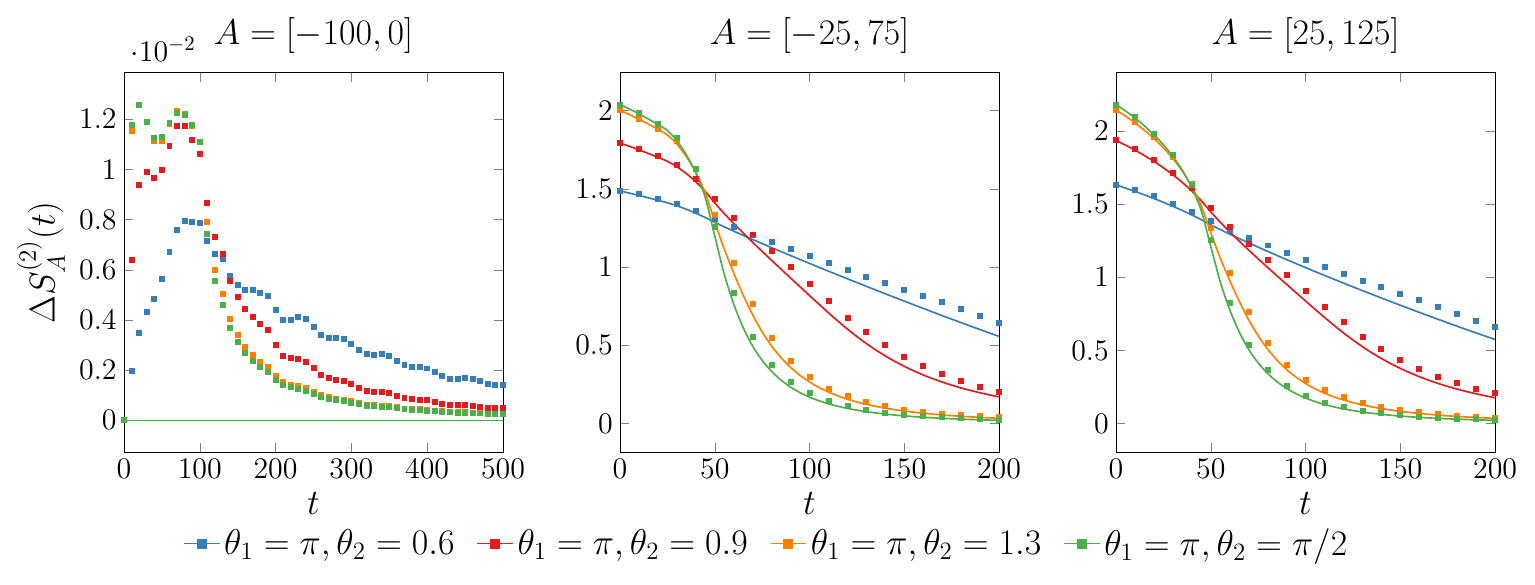}
    \caption{R\'enyi-2 entanglement asymmetry \(\Delta S^{(2)}_A(t)\) as a function of time for several initial angles \(\theta_1, \theta_2\). Symbols correspond to exact  asymmetry obtained from the correlation matrix through Eqs.~\eqref{eq:z_n} and \eqref{eq:entanglement_asymmetry}, taking $L=1600$ as a chain cutoff,  while solid curves correspond to Eq. \eqref{eq:entanglement_asymmetry} with the GHD prediction~\eqref{eq:deltas2_ghd} for the charged moments. In all plots, the  subsystem length $\ell=100$. In the lower panels, all the spins in the left half of the chain are fully polarized in the $z$ direction. The left, middle, and right columns correspond to subsystems located entirely within the left half, crossing the interface, and entirely within the right half, respectively. }
    \label{fig:entanglement_asymmetry2}
\end{figure}

\begin{figure}[t!]
    \centering
    \includegraphics[width=\textwidth]{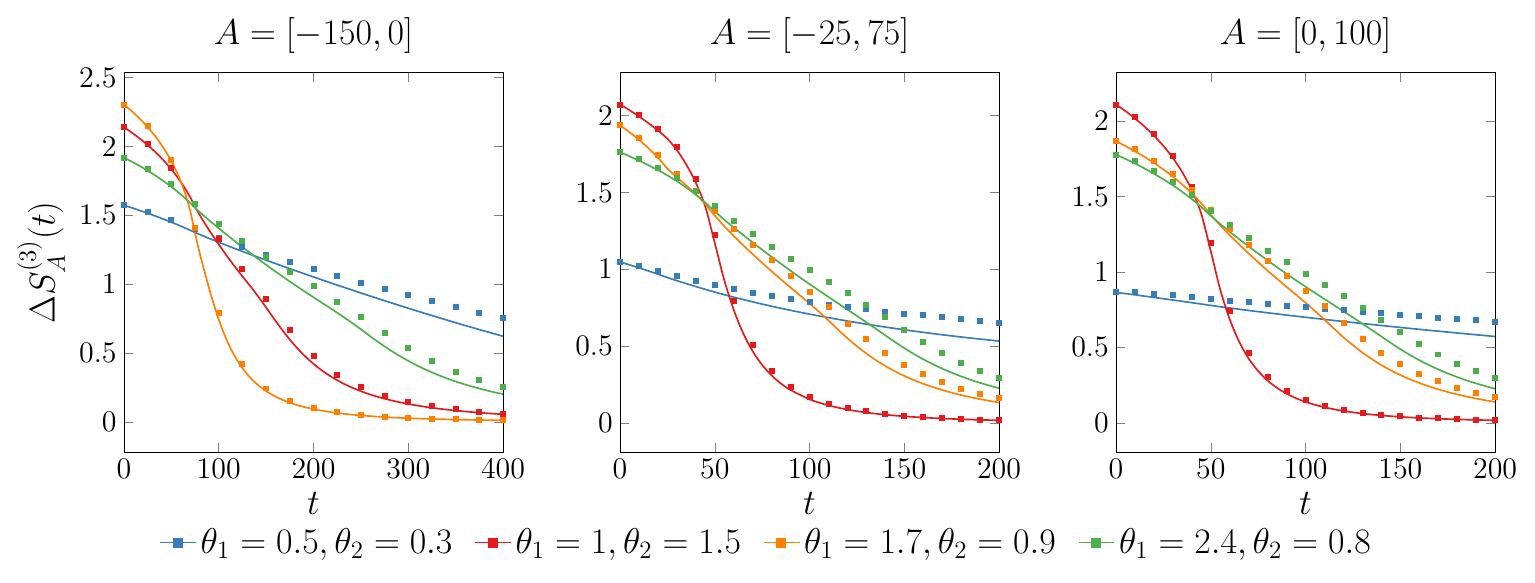}
     \includegraphics[width=\textwidth]{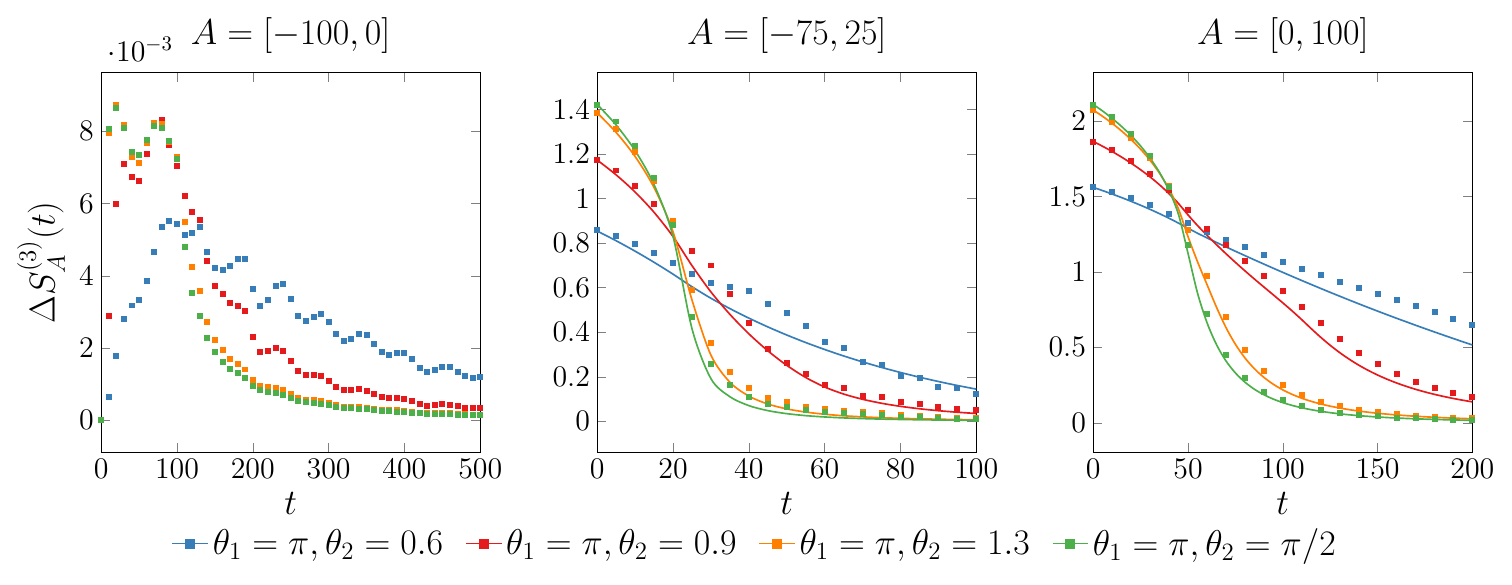}
    \caption{R\'enyi-3 entanglement asymmetry $\Delta S^{(3)}_A(t)$ as a function of time for several initial angles $\theta_1, \theta_2$. Symbols correspond to exact  asymmetry obtained from the correlation matrix through Eqs.~\eqref{eq:z_n} and \eqref{eq:entanglement_asymmetry}, using a numerical cutoff $L=1600$, while solid curves correspond to Eq. \eqref{eq:entanglement_asymmetry} with the GHD prediction~\eqref{eq:deltas2_ghd} for the charged moments. The subsystem lengths are $\ell=150$ in the upper left panel and $\ell=100$ for the rest of panels. In the lower row,  all the spins in the left half of the chain are fully polarized in the $z$ direction. The left, middle, and right columns correspond to subsystems located entirely within the left half, crossing the interface, and entirely within the right half, respectively. }
    \label{fig:entanglement_asymmetry3}
\end{figure}

In Figs.~\ref{fig:entanglement_asymmetry2} and \ref{fig:entanglement_asymmetry3}, we benchmark the GHD prediction against the exact R\'enyi entanglement asymmetry for R\'enyi indices $n=2$ and $3$, respectively. The charged moments entering the definition of the asymmetry are computed using either the exact expression~\eqref{eq:z_n} (symbols) or its GHD approximation~\eqref{eq:deltas2_ghd} (solid lines).
As in the entanglement entropy figures, the left, middle, and right panels correspond to subsystems located entirely within the left half of the chain, straddling the interface, and entirely within the right half, respectively. In the lower row of both figures, the left half of the chain is initially fully polarized down ($\theta_1=\pi$). Let us discuss these results considering separately the different positions of the subsystem relative to the interface.

When the subsystem initially breaks the $U(1)$ symmetry, the asymmetry starts from a nonzero value, as expected. It eventually decays toward zero at long times, reflecting the dynamical restoration of the $U(1)$ symmetry within the subsystem. For these subsystems, the GHD prediction describes the exact evolution of the asymmetry well, although discrepancies arise when the initial state approaches a fully polarized configuration (e.g. the initial state $(\theta_1,\theta_2)=(0.5,0.3)$). The accuracy of the GHD approximation improves for intermediate angles $\theta_1,\theta_2$ and with increasing subsystem size.

\textbf{Subsystem within a tilted region.} Within the quasiparticle picture, only quasiparticle pairs originating inside the subsystem contribute to the charged moments.  Consequently, when the subsystem lies entirely within one half of the chain with $\theta_j\not\in\{0, \pi\}$, the GHD prediction~\eqref{eq:deltas2_ghd} for the charged moments coincides with that of the corresponding homogeneous quench, independently of the tilt angle of the other half.  In particular, we recover the quantum Mpemba effect previously observed for homogeneous tilted ferromagnetic initial states~\cite{entanglement_asymmetry}, whereby states with larger initial asymmetry can restore the symmetry faster than those with smaller initial asymmetry. The microscopic mechanism underlying this effect is related to the momentum distribution of the symmetry-breaking Cooper pairs~\cite{rkacmb-24, xy}. Faster pairs cease to contribute to the asymmetry as their members separate and leave the subsystem, whereas the long-time behavior is dominated by the slowest pairs, with momenta near $k=0$ and $\pi$ in our case. For homogeneous tilted ferromagnetic states, the initial asymmetry increases with the tilt angle $\theta_j$ towards $\pi/2$, whereas the contribution of the slowest pairs decreases. Consequently, the quantum Mpemba effect is guaranteed for any two distinct tilt angles $0<\theta_j\leq \pi/2$ in our inhomogeneous quench, provided the subsystem lies entirely within one half of the chain.

\textbf{Subsystem crossing the interface.} 
The results presented in Figs.~\ref{fig:entanglement_asymmetry2} and \ref{fig:entanglement_asymmetry3}  show that the quantum Mpemba effect can also occur when the subsystem crosses the interface. However, the single crossing observed in these examples is not the only possible behavior. In this configuration, quasiparticle pairs originating in both halves contribute to the charged moments, with each contribution depending on the corresponding initial tilt angle and the portion of the subsystem in which the pairs originate. The competition between these contributions can lead to a richer symmetry-restoration dynamics. To understand this behavior, we examine the long-time behavior of the R\'enyi-2 entanglement asymmetry. 
For generic tilt angles $0<\theta_1,\theta_2<\pi$, expanding Eq.~\eqref{eq:deltas2_ghd} around the slow modes $k=0$ and $\pi$ yields, in the regime $t\gg\ell$ (see Appendix~\ref{app}),
\begin{equation}\label{eq:large_t_ea}
\Delta S_A^{(2)}(t)\simeq
\frac{\ell^4}{192\pi t^3}
\left[
w_1g(\theta_1)+w_2g(\theta_2)
\right],
\end{equation}
where
\begin{equation}\label{eq:g_theta}
g(\theta)=1+\frac{8\cos^2\theta}{\sin^4\theta},
\end{equation}
and $w_1,w_2$ are geometrical weights satisfying $w_1+w_2=1$.
The asymptotic expansion is not uniform in the limits $\theta\to0,\pi$. In particular, the long-time limit and the fully polarized limits do not commute, so Eq. \eqref{eq:g_theta} cannot be directly applied to configurations containing fully polarized regions.
Denoting by $d_1$ and $d_2$ the lengths of the portions of the subsystem in the left and right halves, respectively, with $d_1+d_2=\ell$, these weights are given by
\begin{equation}
w_1=
\begin{cases}
8(d_1/\ell)^4, & d_1\leq\ell/2,\\
1-8(d_2/\ell)^4, & d_1>\ell/2,
\end{cases}
\qquad w_2=1-w_1.
\end{equation}
In particular, for a subsystem centered on the interface, $w_1=w_2=1/2$. On the other hand, for sufficiently large subsystems, the initial R\'enyi-2 entanglement asymmetry behaves as (see Appendix~\ref{app})
\begin{equation}\label{eq:t_0_ea}
\Delta S_A^{(2)}(0)\simeq
\frac{1}{2}\log\left[
d_1\sin^2\theta_1+d_2\sin^2\theta_2
\right]+\frac{1}{2}\log\frac{\pi}{4}.
\end{equation}
Therefore, for two initial configurations ${\rm I}$ and ${\rm II}$, an asymptotic reversal of the ordering of their entanglement asymmetries, characteristic of the quantum Mpemba effect, occurs when
\begin{equation}
\begin{aligned}
d_1\sin^2\theta_1^{\rm I}+d_2\sin^2\theta_2^{\rm I}
&>
d_1\sin^2\theta_1^{\rm II}+d_2\sin^2\theta_2^{\rm II},\\
w_1g(\theta_1^{\rm I})+w_2g(\theta_2^{\rm I})
&<
w_1g(\theta_1^{\rm II})+w_2g(\theta_2^{\rm II}).
\end{aligned}
\label{eq:mpemba_cond}
\end{equation}
When the subsystem lies entirely within one half of the chain, the corresponding geometrical weight equals one, while the other vanishes, recovering the homogeneous result for the tilt angle of that half~\cite{entanglement_asymmetry,xy}.

The conditions~\eqref{eq:mpemba_cond} show that, unlike the homogeneous case or when the subsystem lies entirely within one half of the chain, the quantum Mpemba effect is not guaranteed for arbitrary pairs of initial tilt angles when the subsystem crosses the interface. For tilt angles $0<\theta_j<\pi/2$, if $\theta_1^{\rm I}>\theta_1^{\rm II}$ and $\theta_2^{\rm I}>\theta_2^{\rm II}$, the Mpemba effect is guaranteed, since the initial asymmetry increases with both tilt angles, whereas $g(\theta)$ decreases monotonically as $\theta$ approaches $\pi/2$. However, if $\theta_1^{\rm I}>\theta_1^{\rm II}$ and $\theta_2^{\rm I}<\theta_2^{\rm II}$ (or vice versa), the occurrence of the effect depends on the competition between the contributions from the two halves and on how much of the subsystem lies within each half. In particular, a configuration with a larger initial asymmetry can also have a larger long-time asymmetry, preventing an asymptotic reversal of their ordering and the occurrence of the quantum Mpemba effect.

\begin{figure}[t!]
    \centering
    \includegraphics[width = 0.45\textwidth]{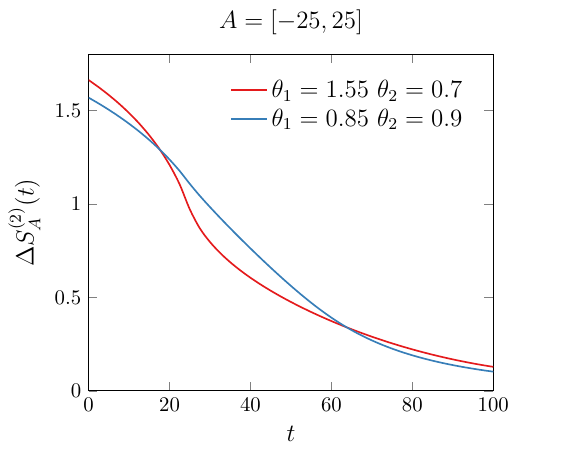}
     \includegraphics[width = 0.45\textwidth]{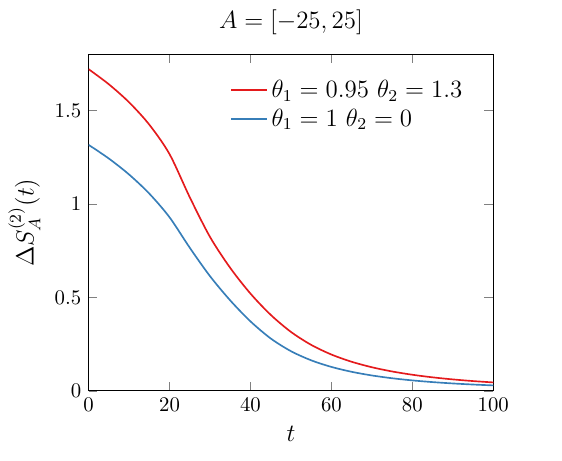}
    \caption{Examples of the behavior of the R\'enyi-2 entanglement asymmetry for subsystems crossing the interface. We compare two pairs of initial ferromagnetic configurations for a subsystem of length $\ell=50$ centered at the interface of the chain. The curves correspond to the GHD prediction. While the asymmetry curves cross twice in the left panel and do not cross in the right panel (within the time window considered), in both cases the configuration with the larger initial asymmetry also has the larger long-time asymmetry, so no Mpemba effect occurs.}
    \label{fig:mult_no_crossings}
\end{figure}

Interestingly, the absence of a Mpemba effect does not exclude crossings of the entanglement asymmetries at intermediate times. For example, in Fig.~\ref{fig:mult_no_crossings} left, we consider two initial configurations ${\rm I}$ and ${\rm II}$ with $(\theta_1^{\rm I},\theta_2^{\rm I})=(1.55,0.7)$ and $(\theta_1^{\rm II},\theta_2^{\rm II})=(0.85,0.9)$, respectively. For a subsystem centered on the interface, the GHD prediction exhibits two crossings: configuration ${\rm I}$ has a larger asymmetry initially and at long times, but a smaller asymmetry at intermediate times. By contrast, for $(\theta_1^{\rm I},\theta_2^{\rm I})=(0.95,1.30)$ and $(\theta_1^{\rm II},\theta_2^{\rm II})=(1.00,0)$, no crossings are observed within the time window considered, as shown in Fig.~\ref{fig:mult_no_crossings} right. These examples illustrate how the competition between the contributions from the two halves can give rise to different symmetry-restoration dynamics, including multiple crossings~\cite{carc-24} and the absence of the Mpemba effect.

\textbf{Subsystem within a non-tilted region.}
Remarkably, the situation is different when the subsystem lies entirely within the initially $z$-polarized left half of the chain, with all spins aligned along the negative $z$ direction  (bottom-left panels of Figs.~\ref{fig:entanglement_asymmetry2} and~\ref{fig:entanglement_asymmetry3}). Unlike the other subsystem configurations, here the $U(1)$ symmetry is initially respected within the subsystem, and consequently the entanglement asymmetry vanishes. According to the GHD prediction~\eqref{eq:deltas2_ghd}, the asymmetry remains identically zero at all times. To understand this result, recall that, within the Euler-scale quasiparticle picture, the leading contribution to the entanglement asymmetry comes from Cooper pairs whose two members are contained within the subsystem. However, no Cooper pairs originate in the fully polarized half, while those originating in the tilted half propagate in opposite directions and can never both be contained within a subsystem located entirely in the left half. Nevertheless, as shown in the figures, the exact asymmetry becomes nonzero during the evolution, although it remains small (of order $10^{-2}$), indicating the presence of contributions beyond the leading-order GHD description. The asymmetry initially grows, then oscillates around a plateau, and eventually decreases toward zero at long times. We have numerically verified that all these features, including the height of the plateau, remain quantitatively unchanged upon increasing the numerical cutoff $L$.

%\begin{figure}[t!]
%    \centering
%    \includegraphics[width=\textwidth]{Figures/Entanglement_asymmetry_order3_initially_polarized_fig10.pdf}
%    \caption{Entanglement asymmetry \(\Delta S^{(3)}_A(t)\) as a function of time for several initial angles \(\theta_1 = \pi, \theta_2\). Here, the left region of the chain is initially down polarized. Symbols correspond to exact correlation-matrix results [Eqs.~\eqref{eq:z_n} and \eqref{eq:entanglement_asymmetry}], while solid lines represent the GHD prediction [Eqs.~\eqref{eq:deltas2_ghd} and \eqref{eq:entanglement_asymmetry}]. Here, the left part of the chain is initially fully polarized down. Geometries and parameters are identical to Fig.~\ref{fig:several_theta3_polarized}.}
    %\label{fig:entanglement_asymmetry3_initially_polarized}
%\end{figure}

\begin{figure}[t!]
    \centering
    \includegraphics[width = \textwidth]{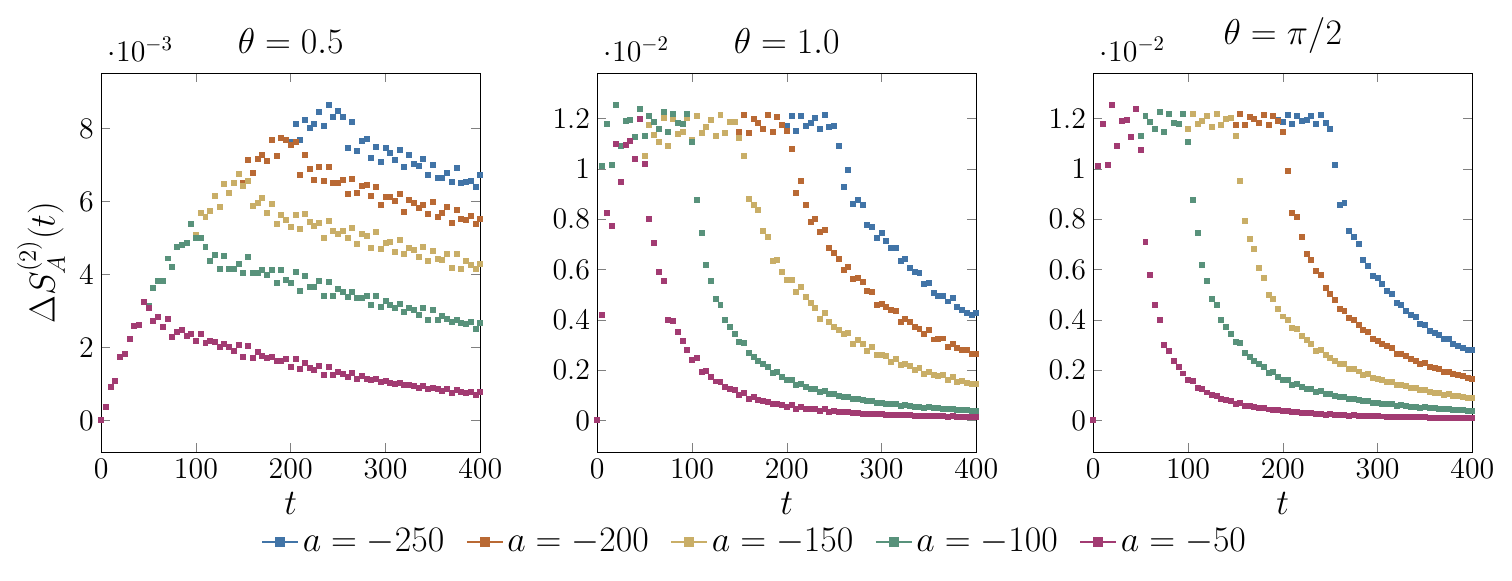}
    \caption{Time evolution of R\'enyi-2 entanglement asymmetry for a chain initialized with all spins in the left half  fully polarized down along the $z$-axis, while those in the right half are tilted by an angle $\theta$. We report several subsystems $A=[a,b]$ entirely contained in the left half of the chain. The right subsystem endpoint is fixed at $b=0$, while the subsystem size is varied by changing the left endpoint $a$. Different panels correspond to different values of the initial tilt angle $\theta$ in the right half. Symbols represent the exact correlation-matrix results obtained using Eq.~\eqref{eq:entanglement_asymmetry}, taking as a chain cutoff $L=1600$. For these  configurations, the GHD formula~\eqref{eq:deltas2_ghd} predicts an identically vanishing entanglement asymmetry at all times.
}
    \label{fig:entanglement_entropy_left_severala}
\end{figure}

In Fig.~\ref{fig:entanglement_entropy_left_severala}, we examine in more detail the evolution of the asymmetry for a subsystem located entirely within the fully polarized half of the chain, focusing on its dependence on the initial tilt angle $\theta_2$ of the other half and the subsystem length $\ell$.
%Although it is difficult to accurately describe the exact behavior of the system in the left part beyond the GHD approximation, we can nevertheless highlight some remarks and interesting features.
Each panel corresponds to a different tilt angle $\theta_2$ in the right half of the chain, and for each angle, we consider several sizes of the subsystem $A$, whose right endpoint is always at site $b=0$, adjacent to the interface between the two halves. For small values of $\theta_2$ (left panel), $\Delta S_A^{(2)}(t)$ initially grows approximately logarithmically, reaches a maximum that increases with the subsystem size, and then decreases without developing a plateau. As $\theta_2$ increases, the initial growth becomes steeper, while the numerical results for different subsystem sizes nearly coincide in this regime. This initial growth can be associated with the propagation of correlations that break the $U(1)$ symmetry from the tilted right half into the initially polarized subsystem. For sufficiently large $\theta_2$ (middle and right panels), the asymmetry reaches an approximately constant value, $\Delta S_A^{(2)}\simeq 1.2\times 10^{-2}$, seemingly independent of both $\theta_2$ and the subsystem size, giving rise to a plateau before the subsequent decay. Interestingly, the decay begins at times of order $t\sim\ell$ in all three panels, consistent with the timescale for symmetry-breaking correlations to propagate from the interface to the left boundary of the subsystem. For smaller $\theta_2$, the growth is slower, and the decay sets in before the asymmetry reaches its apparent saturation value. The nearly constant plateau height suggests a saturation of the symmetry-breaking correlations generated across the interface. %Interestingly, this saturation value appears to be independent of both $\theta_2$ and the subsystem size. 
This apparent saturation seems to be specific to the present quench protocol and is much smaller than the maximum asymmetry attainable within the class of Gaussian states, which scales as $\frac{1}{2}\log\ell$ for large subsystems~\cite{ammscp-26} (see also Refs.~\cite{mcp-26, gam-26}). 
Its microscopic origin, however, remains unclear and may require incorporating quantum fluctuations beyond the Euler-scale GHD description or be generated by other subleading effects.

\section{Conclusions}
\label{sec:conclusions}

In this work, we have investigated the dynamics of entanglement asymmetry following an inhomogeneous quantum quench in the spin-$1/2$ XX chain, with the two halves initially prepared in tilted ferromagnetic states with different tilt angles. We have extended the quasiparticle picture underlying the entanglement asymmetry, originally developed for homogeneous symmetry-breaking initial states, to inhomogeneous settings by combining it with generalized hydrodynamics. Within generalized hydrodynamics, the state of the system is locally described at large space-time scales by a quasiparticle occupation function that evolves ballistically and accounts for the spatial inhomogeneity of the initial state. We combine this hydrodynamic description with the quasiparticle picture of charged moments, in which pairs of oppositely propagating quasiparticles, interpreted as Cooper pairs, carry the pairing correlations responsible for the initial $U(1)$ symmetry breaking. Combining these ingredients, we derive Euler-scale predictions for the R\'enyi entanglement entropies and charged moments, from which the entanglement asymmetry is obtained. Comparison with exact numerical results from the time-evolved two-point fermionic correlation matrices shows good agreement for generic initial tilt angles, while discrepancies become more pronounced when one of the initially prepared regions is fully polarized and hence locally symmetric.

Our results show that the initial inhomogeneity and the position of the subsystem play a crucial role in the dynamics of the asymmetry. For subsystems in which the $U(1)$ symmetry is initially broken, the entanglement asymmetry starts from a nonzero value and generally decreases monotonically toward zero at long times. Moreover, we find that the quantum Mpemba effect persists in spatially inhomogeneous quenches, but its occurrence depends sensitively on the initial tilt angles and the subsystem position. In particular, subsystems crossing the interface can exhibit a richer dynamics, including multiple crossings of the entanglement asymmetry or the absence of an asymptotic reversal of its ordering.

A particularly interesting behavior emerges when the subsystem lies entirely within an initially fully polarized half of the chain. In this configuration, the Euler-scale GHD prediction for the entanglement asymmetry vanishes identically, whereas the exact asymmetry becomes nonzero during the evolution. This reveals the presence of symmetry-breaking correlations propagating from the tilted half into the initially polarized region whose contribution is not captured by the leading-order quasiparticle picture. We have found that the asymmetry initially grows and eventually decreases toward zero, possibly developing a plateau when it reaches an apparent saturation value. Interestingly, this saturation value seems to be independent of both the initial tilt angle of the other half of the chain and the subsystem size, while the onset of the decay occurs at times of order the subsystem length. Understanding the microscopic origin of this behavior may require incorporating quantum fluctuations beyond the Euler-scale GHD description. This behavior can also be interpreted from the perspective of quantum resource theories, where symmetry breaking constitutes a resource and symmetry-preserving operations are \textit{free}~\cite{summer-26, brs-07, vawj-08, gms-09, ms-14, cg-19, kusuki-26, asst-26}. In our setup, this resource is initially confined to the tilted half of the chain, while the fully polarized half is resource-free. The subsequent evolution, although preserving the $U(1)$ symmetry, allows symmetry-breaking correlations to spread into the initially symmetric region. Similar spreading of spatially localized quantum resources under free dynamics has recently been investigated in quantum circuits~\cite{ats-26}. Interestingly, in our case, the resulting local asymmetry remains small and appears to saturate at a characteristic value, suggesting that the redistribution of this resource is constrained by the dynamics of the quench. 

Overall, our results extend the study of entanglement asymmetry and the quantum Mpemba effect to spatially inhomogeneous initial states, while highlighting the limitations of the leading-order hydrodynamic description in capturing symmetry-breaking correlations in certain subsystem configurations. The same framework could also be applied to other spatially inhomogeneous initial states that can be described within GHD. 
Here, we restricted to infinite chains, but one could also consider finite chains and study the finite-size effects, such as revivals. These could in principle be incorporated into the GHD predictions~\cite{asw-22} and in the quasiparticle picture~\cite{mac-20}.
The same is true for the case of disjoint subsystems, whose treatment should be amenable to a similar quasiparticle analysis \cite{ac-19,ac-18b}.

A natural direction for future work is to extend our analysis to interacting integrable systems~\cite{abf-19, tckb-26}. Although we have focused here on a free-fermion model, the quasiparticle picture for the entanglement asymmetry can also be formulated in interacting integrable systems by incorporating the  dressing of quasiparticle properties~\cite{rkacmb-24, bkccr-24, rvc-24, klobas-24}. This would allow us to investigate how interactions affect local symmetry restoration and the quantum Mpemba effect in spatially inhomogeneous quenches. Other possible extensions include introducing point defects at the junction~\cite{ep-12, fg-21, csrc-23}, time-dependent random hopping amplitudes~\cite{raca-26, raca-26-2}, or inhomogeneous potentials~\cite{cvcg-23, ebs-26}. These modifications would allow us to explore how scattering and spatial or temporal variations in the Hamiltonian affect the propagation of symmetry-breaking correlations and the dynamics of entanglement asymmetry. Furthermore, since the dynamics of entanglement asymmetry has also been investigated in nonintegrable systems, including quantum circuits~\cite{liu-24, tcdl-25, yu-25, amt-26, arst-26, yac-26, fcb-25,li-26} and spin chains~\cite{yjzxf-25, ya-26, mpf-26, lzy-25}, it would be interesting to extend these studies to spatially inhomogeneous initial states.
Finally, while we have focused here on \(U(1)\) symmetry, entanglement asymmetry has also been investigated for more exotic symmetries, including discrete \cite{fac-23b}, non-Abelian \cite{rac-24}, gauge \cite{chmp-21,Casini_2022,fm-26} and non-invertible \cite{bcf-25,bgv-26,vca-26,lgh-26} ones. It would therefore be interesting to explore how spatial inhomogeneities affect the dynamics of symmetry restoration and the possible emergence of the quantum Mpemba effect in these more general settings.
\\

\textbf{Acknowledgements.} All authors acknowledge support from the European Research Council under the Advanced Grant no. 101199196 (MOSE). F.A. acknowledges support from the grant RYC2024-050650-I, funded by MICIU/AEI/10.13039/501100011033 and FSE+, from the grants PID2024-160228NB-I00, funded by MICIU/AEI/10.13039/501100011033 and ERDF - A way of making Europe, and DGA-FSE 2023-E21-26R, funded by the Government of Arag\'on, Spain, and from the European Union NextGeneration EU Recovery and Resilience Program on Astrof\'isica y F\'isica de Altas Energ\'ias CEFCA-CAPA-ITAINNOVA.

\textbf{AI statement}:
ChatGPT 5.6 has been used for text polishing and proofreading of the manuscript, including suggestions for improving clarity and grammar. The authors reviewed and verified all generated content and remain fully responsible for the accuracy, interpretation, and presentation of the results.

\appendix

\section{Asymptotic behavior of the entanglement asymmetry}\label{app}

In this appendix, we derive the asymptotic expressions for the R\'enyi-2 entanglement asymmetry reported in Eqs.~\eqref{eq:large_t_ea} and~\eqref{eq:t_0_ea}. Following the approach used for homogeneous quenches in Ref.~\cite{xy}, we analyze the initial asymmetry through a saddle-point approximation and the long-time behavior through an expansion around the slowest quasiparticle modes.

We consider a subsystem $A=[a,b]$ crossing the interface between the two initially tilted halves. For convenience, we place the interface at $x=0$ and define
\begin{equation}
a=-d_1,\quad b=d_2,\quad \ell=d_1+d_2.
\end{equation}

\textbf{Asymptotic behavior at $t=0$.} At $t=0$, the $n=2$ charged moments~\eqref{eq:deltas2_ghd} simplify to
\begin{equation}\label{eq:app_z2}
\log\frac{Z_2(\alpha,0)}{Z_2(0,0)}
=
\frac{d_1}{2}
\int_0^{2\pi}\frac{dk}{2\pi}\,
\log\left[
1-\sin^2\Delta_k(\theta_1)\sin^2\alpha
\right]+\frac{d_2}{2}
\int_0^{2\pi}\frac{dk}{2\pi}\,
\log\left[
1-\sin^2\Delta_k(\theta_2)\sin^2\alpha
\right].
\end{equation}
That is, the charged moments separate into two contributions, each corresponding to the charged moments of a subsystem of length $d_j$ in a homogeneous tilted ferromagnetic state with angle $\theta_j$~\cite{entanglement_asymmetry}.

Inserting~\eqref{eq:app_z2} into the expression for the entanglement asymmetry in Eq.~\eqref{eq:entanglement_asymmetry}, for sufficiently large $d_1$ and $d_2$, the integral over $\alpha$ is dominated by the neighborhoods of $\alpha=0$ and $\alpha=\pi$, where $Z_2(\alpha,0)/Z_2(0,0)=1$. Expanding around $\alpha=0$, we obtain
\begin{equation}
\log\frac{Z_2(\alpha,0)}{Z_2(0,0)}
\simeq
-\frac{\alpha^2}{2}
\left[d_1
\int_0^{2\pi}\frac{dk}{2\pi}\sin^2\Delta_k(\theta_1)
+d_2
\int_0^{2\pi}\frac{dk}{2\pi}\sin^2\Delta_k(\theta_2)
\right].
\end{equation}
Taking into account that
\begin{equation}
\int_0^{2\pi}\frac{dk}{2\pi}\sin^2\Delta_k(\theta_j)=\frac{\sin^2\theta_j}{2},
\end{equation}
we have
\begin{equation}
\log\frac{Z_2(\alpha,0)}{Z_2(0,0)}
\simeq -\frac{\alpha^2}{4}\left(d_1\sin^2\theta_1+d_2\sin^2\theta_2\right).
\end{equation}
The second saddle point, at $\alpha=\pi$, gives an identical contribution. Accounting for both saddle points, we find
\begin{equation}
\Delta S_A^{(2)}(0)\simeq -\log\left[2\int_{-\infty}^\infty \frac{d\alpha}{2\pi}e^{-\frac{\alpha^2}{4}\left(d_1\sin^2\theta_1+d_2\sin^2\theta_2\right)}\right].
\end{equation}
Applying the standard formulas for Gaussian integrals, we finally obtain Eq.~\eqref{eq:t_0_ea} of the main text. 

\textbf{Asymptotic behavior at long times.} We now consider the regime $t\gg\ell$. According to Eq.~\eqref{eq:deltas2_ghd}, only the quasiparticles satisfying $
|v(k)|<\ell/(2t)$ contribute to the charged moments. In particular, since $v(k)=\sin k$, only the momenta close to $k=0$ and $k=\pi$, where the quasiparticle velocity vanishes, contribute in the $t\gg\ell$ limit. 
The difference from the homogeneous case arises from the spatial integration over the emission points. For both quasiparticles to remain inside the subsystem $A=[-d_1,d_2]$ at time $t$, their emission point must satisfy
$-d_1+|v(k)|t\leq y\leq d_2-|v(k)|t$, with $0\leq |v(k)|t\leq\ell/2$.
Let us denote $\ell_j(|v(k)|t)$ the length of the interval $[-d_1+|v(k)|t,d_2-|v(k)|t]$ lying within the half $j$, 
\begin{equation}
\begin{aligned}
\ell_1(r)&=\max\{0,\min(d_1-r,\ell-2r)\},\\
\ell_2(r)&=\max\{0,\min(d_2-r,\ell-2r)\},
\end{aligned}
\qquad 0\leq r\leq\ell/2.
\label{eq:l_j}
\end{equation}
We can rewrite the charged moments~\eqref{eq:deltas2_ghd} for $n=2$ in terms of these lengths as
\begin{equation}\label{eq:app_z2_2}
\log\frac{Z_2(\alpha,t)}{Z_2(0,t)}
=\frac12\sum_{j=1}^{2}
\int_0^{2\pi}\frac{dk}{2\pi}\,
\Theta\bigl(\ell-2|v(k)|t\bigr)
\mathcal \ell_j(|v(k)|t)
\log\left[1-\sin^2\Delta_k(\theta_j)\sin^2\alpha\right].
\end{equation}
Therefore, each half contributes as in the corresponding homogeneous tilted ferromagnetic state, with its contribution weighted by the length $\ell_j(r)$ of the interval $[-d_1+r,d_2-r]$ within that half.

For the modes around $k=0, \pi$,  $\sin^2\Delta_k(\theta_j)$ is small and behaves as 
\begin{equation}
\sin^2\Delta_k(\theta_j)\simeq
\begin{cases}
\displaystyle
\frac{\gamma^2(\theta_j)}{[1+h(\theta_j)]^2}k^2,
& k\simeq0,\\[3mm]
\displaystyle
\frac{\gamma^2(\theta_j)}{[1-h(\theta_j)]^2}(k-\pi)^2,
& k\simeq\pi.
\end{cases}
\end{equation}
Therefore, we can expand the logarithm in Eq.~\eqref{eq:app_z2_2} to leading order $\sin^2\Delta_k(\theta_j)$, obtaining
\begin{equation}
\log\frac{Z_2(\alpha,t)}{Z_2(0,t)}
\simeq
-\frac{\sin^2\alpha}{2}
\sum_{j=1}^{2}
\int_0^{2\pi}\frac{dk}{2\pi}\,
\Theta\bigl(\ell-2|v(k)|t\bigr)
\sin^2\Delta_k(\theta_j)
\ell_j(|v(k)|t).
\end{equation}
We now evaluate the momentum integrals around $k=0,\pi$. Defining $\delta_t=\arcsin(\ell/2t)$, we expand around each slow mode using $q=k-k_*$, with $k_*=0,\pi$, $\sin^2\Delta_k(\theta_j)\simeq c_{k_*}(\theta_j)q^2$, and $|v(k)|\simeq|q|$,
\begin{equation}
\int_{k_*-\delta_t}^{k_*+\delta_t}
\frac{dk}{2\pi}\,
\sin^2\Delta_k(\theta_j)
\ell_j(|v(k)|t)
\simeq
\frac{c_{k_*}(\theta_j)}{2\pi}
\int_{-\ell/(2t)}^{\ell/(2t)}
dq\, q^2\ell_j(t|q|).
\end{equation}
Summing the contributions from the two slow modes and using
\begin{equation}
c_0(\theta_j)+c_\pi(\theta_j)=2g(\theta_j),
\end{equation}
where $g(\theta)$ was introduced in Eq.~\eqref{eq:g_theta}, we find
\begin{equation}
\log\frac{Z_2(\alpha,t)}{Z_2(0,t)}
\simeq
-\frac{\sin^2\alpha}{\pi t^3}
\left[g(\theta_1)J_1+
g(\theta_2)J_2\right],
\end{equation}
where
\begin{equation}\label{eq:J_j}
J_j=\int_0^{\ell/2}dr\,r^2\ell_j(r).
\end{equation}
Inserting this result into the definition of the R\'enyi-2 entanglement asymmetry~\eqref{eq:entanglement_asymmetry} and performing the integral over $\alpha$, we obtain
\begin{equation}
\Delta S_A^{(2)}(t)\simeq \frac{g(\theta_1)J_1+g(\theta_2)J_2}{2\pi t^3}.
\end{equation}
Using Eq.~\eqref{eq:l_j} in~\eqref{eq:J_j}, we find the explicit expression for the geometric weights $J_j$,
\begin{equation}
J_1=
\begin{cases}
\dfrac{d_1^4}{12}, & d_1\leq\ell/2,\\[6pt]
\dfrac{\ell^4}{96}-\dfrac{d_2^4}{12},
& d_1>\ell/2,
\end{cases}
\qquad
J_2=\frac{\ell^4}{96}-J_1.
\end{equation}
Normalizing them as 
\begin{equation}
w_j=\frac{J_j}{J_1+J_2},\quad w_1+w_2=1,
\end{equation}
we finally arrive at Eq.~\eqref{eq:large_t_ea} of the main text.

\end{document}